\documentclass[twocolumn]{aastex63}

\shorttitle{A Low Incidence of Mid-Infrared Variability in Dwarf Galaxies}
\shortauthors{Secrest \& Satyapal}

\begin{document}

\title{A Low Incidence of Mid-Infrared Variability in Dwarf Galaxies}

\correspondingauthor{Nathan Secrest}
\email{nathan.secrest@navy.mil}

\author[0000-0002-4902-8077]{Nathan J. Secrest}
\affiliation{U.S. Naval Observatory \\
3450 Massachusetts Ave NW \\
Washington, DC 20392-5420, USA}

\author[0000-0003-2277-2354]{Shobita Satyapal}
\affiliation{George Mason University \\
4400 University Dr \\
Fairfax, VA 22030-4444, USA}

\begin{abstract}
Using 8.4 years of photometry from the AllWISE/NEOWISE multi-epoch catalogs, we compare the mid-infrared variability properties of a sample of 2197 dwarf galaxies ($M_\star<2\times10^9$~$h^{-2}M_\sun$) to a sample of 6591 more massive galaxies ($M_\star\geq10^{10}$~$h^{-2}M_\sun$) matched in mid-infrared apparent magnitude. We find only 2 dwarf galaxies with mid-infrared variability, a factor of $\sim10$ less frequent than the more massive galaxies ($p=6\times10^{-6}$), consistent with previous findings of optical variability in low-mass and dwarf galaxies using data with a similar baseline and cadence. Within the more massive control galaxy population, we see no evidence for a stellar mass dependence of mid-infrared variability, suggesting that this apparent reduction in the frequency of variable objects occurs below a stellar mass of $\sim10^{10}$~$h^{-2}M_\sun$. Compared to the more massive galaxies, AGNs selected in dwarf galaxies using either their mid-infrared color or optical emission line classification are systematically missed by variability selection. Our results suggest, in agreement with previous optical studies at similar cadence, that variability selection of AGNs in dwarf galaxies is ineffective unless higher-cadence data is used.
\end{abstract}


\section{Introduction} \label{section: intro}
Over the past several decades, it has become clear that a large fraction of active galactic nuclei (AGNs) are missed by optical spectroscopic, mid-infrared, X-ray, and radio surveys due either to obscuration of the central engine, or dilution of the AGN emission from star formation in the host galaxy \citep[e.g.,][]{goulding2009,2015ApJ...811...26T,satyapal2018}. This is a significant deficiency, particularly in dwarf galaxies, since the massive black hole (MBH:~$M_\mathrm{BH}>10^2$~$M_\sun$) occupation fraction and mass distribution in the low galaxy mass regime may place important constraints on models of supermassive black hole ($M_\mathrm{BH}>10^6$~$M_\sun$) ``seed'' formation \citep[e.g.,][]{Volonteri2009, Volonteri2010, vanwassenhove2010, greene2012}. Moreover, given the expected mass range of MBHs in dwarf galaxies ($\sim10^4$M$_{\odot}$ - $10^6$M$_{\odot}$), mergers between MBHs in dwarf galaxies may be detected out to high-redshift by the Laser Interferometer Space Antenna \citep[LISA;][]{amaroseoane2012,amaroseoane2013}. The identification of AGNs in dwarf galaxies is especially challenging because the AGN luminosity is typically low compared with the star formation rate, diluting optical line emission from the AGN narrow line region with nebular emission near starbursts \citep[e.g.,][]{2015ApJ...811...26T}. X-ray AGN selection is similarly impacted, as the X-ray luminosities from AGNs powered by MBHs are often comparable to expectations from high-mass X-ray binaries (HXMBs) and ultraluminous X-ray sources (ULXs), which generally have X-ray luminosities below $10^{41}$~erg~s$^{-1}$ \citep[e.g.,][]{2012MNRAS.419.2095M, 2017ARA&A..55..303K}. 

Mid-infrared AGN selection, which since the advent of the \textit{Wide-field Infrared Survey Explorer} \citep[\textit{WISE};][]{2010AJ....140.1868W} has allowed for the identification of millions of AGNs \citep{2015ApJS..221...12S, 2018ApJS..234...23A,2019MNRAS.485.4539J,2019MNRAS.489.4741S} has shown potential for finding AGNs in the low stellar mass regime \citep[e.g.,][]{2014ApJ...784..113S,2015ApJ...798...38S,2015MNRAS.454.3722S,2019MNRAS.489L..12K}. However, extreme starbursts may mimic AGN activity in the mid-infrared \citep[e.g.,][]{2016ApJ...832..119H}, a consequence of high ionization parameter $U$ and/or gas column density $N_\mathrm{H}$ \citep{satyapal2018}, leading to discrepant results when using mid-infrared color diagnostics to select AGNs in dwarf galaxies \citep[e.g.,][]{2020MNRAS.492.2528L}. The low gas-phase metallicities of dwarf galaxies may also complicate AGN selection, as emission line ratios are sensitive to metallicity and an AGN in a low-metallicity environment may have the same emission line ratios as a low-metallicity starburst \citep[e.g.,][]{groves2006,cann2019}. Low metallicity may also lead to significant enhancement in the HMXB population \citep[e.g.,][]{2014MNRAS.441.2346B}, complicating efforts to use a star formation rate-dependent AGN X-ray luminosity selection cut based on relations found for larger galaxies \citep[e.g., the relation of][]{2010ApJ...724..559L}. Finally, at the lowest black hole masses ($M_\mathrm{BH}\lesssim10^4$~$M_\sun$) the hardening of the accretion disk spectral energy distribution (SED) may preclude optical emission line-based AGN selection altogether \citep{cann2019}.

An additional method to identify AGNs is to search for luminosity variability. AGNs have long been known to be variable at every wavelength, with time scales ranging from hours to years \citep[e.g.,][]{ulrich1997,2007AJ....134.2236S,macleod2010, 2016ApJ...817..119K}. The cause of the variability is possibly related to instabilities in the accretion disk or surface temperature fluctuations \citep[e.g.,][]{ruan2014}. Variability selection is potentially promising for finding MBHs in dwarf galaxies, where the host galaxy emission may dominate the total luminosity \citep[e.g.,][]{2015ApJ...811...26T}. For example, \citet{2010ApJ...716..530K} found that lower luminosity AGNs exhibiting mid-infrared variability in the \textit{Spitzer} Deep Wide-Field Survey \citep{2009ApJ...701..428A} sometimes showed bluer, galaxy-like $[3.6\,\micron]-[4.5\,\micron]$ colors outside of the \citet{2005ApJ...631..163S} AGN selection wedge. Additionally, variability has been found to be stronger in lower luminosity AGNs \citep{1994ApJ...433..494T}. Variability may therefore provide a method of AGN selection free of some of the biases against low-luminosity AGNs inherent to other selection techniques, enabling a more complete AGN census.

Recently, \citet{2018ApJ...868..152B} identified 135 galaxies with AGN-like optical variability using \text{Sloan~Digital~Sky~Survey} \citep[SDSS;][]{2000AJ....120.1579Y} data in the Stripe~82 field. They found that in the dwarf galaxy population, a significant fraction of the variable sources are not identified as AGNs based on their narrow line ratios, demonstrating that variability can reveal AGNs missed by other selection techniques in the low-mass, low-metallicity regime. Using the High~Cadence~Transit~Survey \citep[HiTS][]{2018AJ....156..186M}, \citet{2020ApJ...889..113M}  find hundreds of galaxies with candidate variable AGNs, but the vast majority of them (94.3\%) do not show AGN-like optical emission line ratios, further supporting the use of variability as a means of selecting AGNs otherwise missed by conventional selection techniques.

Mid-infrared variability may offer a new window in which to search for AGNs that elude detection through other selection methods. Since the UV and optical emission produced by the accretion disk is absorbed by surrounding dust and re-radiated in the mid-infrared, flux variations in the accretion disk will produce variability in the mid-infrared, with a time lag corresponding to the distance and geometric distribution of the dust. Mid-infrared variability studies have unique advantages over optical variability studies. They are less sensitive to optically-obscured and Compton-thick AGNs, which are a significant percentage of the nearby low-luminosity AGN population as shown using recent hard X-ray \textit{NuSTAR} observations \citep[e.g.,][]{annuar2015, annuar2017, ricci2016}. Indeed, recent \textit{Spitzer} data reveal that only 23\% of the mid-infrared variable sources are identified by soft X-ray surveys \citep{2018MNRAS.476.1111P}. In addition, because the SEDs of SNe are fainter in the mid-infrared \citep{2016PhDT.......177S}, mid-infrared variability selection is less affected by supernovae contamination than optical variability selection.

With the continued operation of \textit{WISE}, it is now possible to conduct systematic studies of mid-infrared variability in large samples of galaxies. \textit{WISE} initially mapped the sky in four passbands: $W1$ (3.4\,$\mu$m), $W2$ (4.6\,$\mu$m), $W3$ (12\,$\mu$m), and $W4$ (22\,$\mu$m) between 2010~January~7 and 2010~August~6. After depleting the solid hydrogen in its outer cryogen tank, \textit{WISE} continued surveying the sky in $W1$, $W2$, and $W3$, before running out of cryogen completely and continuing onto a post-cryo survey in $W1$ and $W2$ from 2010~September~29 to 2011~February~1 as part of the \textit{Near-Earth Object Wide-field Infrared Survey Explorer} mission \citep[NEOWISE;][]{2011ApJ...731...53M}.\footnote{\url{http://wise2.ipac.caltech.edu/docs/release/allwise/expsup/sec1\_2.html\#phases}} After a 3 year hibernation, \textit{WISE} resumed surveying the sky in $W1$ and $W2$ on 2013~December~23 for the NEOWISE Reactivation mission \citep[NEOWISE-R][]{2014ApJ...792...30M}, resulting in a total baseline of over 8~years. 

In this paper, we conduct the first systematic mid-infrared variability study of dwarf galaxies. The main goal of this paper is to determine the mid-infrared variability characteristics of dwarf galaxies compared to more massive galaxies. In Section~\ref{section: methods}, we describe our sample selection, mid-infrared data, and variability metrics. We discuss our results in Section~\ref{section: results}, which includes a comparison with other AGN selection methods. We give our main conclusions in Section~\ref{section: conclusions}.

\section{Methodology} \label{section: methods}
\subsection{Dwarf and Control Galaxy Samples} \label{subsection: samples}
We produced a sample of galaxies from the NASA-Sloan Atlas (NSA), version \texttt{v1\_0\_1},\footnote{\url{https://www.sdss.org/dr16/manga/manga-target-selection/nsa}} which differs from the original NSA catalog \texttt{v0\_1\_2}\footnote{\url{http://www.nsatlas.org/data}} in that it contains sources out to a redshift of $<0.15$ and improved photometry through the use of elliptical Petrosian magnitudes. The NSA catalog uses photometry from SDSS images produced with the improved sky background subtraction method described in \citet{2011AJ....142...31B}, and includes aperture-matched ultraviolet photometry from the \textit{Galaxy Evolution Explorer} (\textit{GALEX}). Stellar masses in the NSA catalog are fit using the five templates of \citet{2007AJ....133..734B}, which were derived from a basis set of stellar population synthesis models from \citet{2003MNRAS.344.1000B}. NSA catalog stellar masses are widely used in galaxy studies, including in studies of AGNs in dwarf galaxies \citep{2013ApJ...775..116R,2015ApJ...798...38S,2015ApJ...805...12L,2016ApJ...832..119H,2017ApJ...836...20B,2018ApJ...868..152B,2019ApJ...884..180D,2020ApJ...888...36R}.

We cross-matched the entire NSA catalog, which contains 641\,409 entries, to the AllWISE catalog to within the default $10\arcsec$ for completeness, returning 637\,098 unique matches. We created a volume-limited sample of dwarf galaxies by selecting only galaxies with a heliocentric redshift greater than 0.02 to minimize the effect of their peculiar velocities on their distance estimates, and a redshift less than 0.03, below which the sample is unbiased for galaxies with an absolute $r$-band magnitude less than $-17$. We use a stellar mass cut of $M_\star<2\times10^{9}$~~$h^{-2}M_\sun$. In order to compare the incidence and characteristics of mid-infrared variable dwarf galaxies to more massive galaxies, we additionally made a sample of galaxies with a stellar mass greater than $10^{10}$~$h^{-2}$~$M_\sun$. However, in the redshift range of our dwarf galaxy sample the \textit{WISE} photometry of larger galaxies is extended, with 90\% of measurements having a profile-fit photometry reduced $\chi_{W1}^2$ greater than 12.2. To ensure accurate and comparable flux measurements between dwarf galaxies and their controls, we chose control galaxies with redshifts between 0.13 and 0.14, with an $r$-band absolute magnitude cut of $-20.7$. Ninety percent of the measurements in this sample have $\chi_{W1}^2<3.5$. We note that $M_\star<2\times10^{9}$~$h^{-2}M_\sun$ is $\log_{10}(M_\star/M_\sun)<9.6$ using $h=0.73$, as in \citet{2013ApJ...775..116R}, so the dwarf galaxies we study in this work are comparable in stellar mass to those in the \citet{2013ApJ...775..116R} sample.

We retrieved the AllWISE Multiepoch Photometry (AllWISE MEP) Table and the NEOWISE-R Single Exposure (L1b) Source Table data by matching on the coordinates of the AllWISE counterparts to within $3\arcsec$,\footnote{\url{http://wise2.ipac.caltech.edu/docs/release/neowise/expsup/sec2\_1a.html\#allwise\_cntr}} although for the AllWISE MEP sources the offset from the AllWISE catalog is exactly zero for true counterparts. As the individual measurements in the AllWISE MEP and NEOWISE-R are shallower than the AllWISE catalog, detections at faint magnitudes appear brighter than in the AllWISE catalog because of the Eddington bias \citep{2014ApJ...792...30M}, and this bias affects the shallower $W2$ band more than $W1$. For our pool of potential control galaxies, fewer than 5\% of them have AllWISE $W2>14.5$~mag, so we cut both our dwarf galaxy sample and control pool to those above this value. For $W2=14.5$, the maximum deviation from the AllWISE magnitude detectable in the AllWISE MEP and NEOWISE-R data is about $\sim1$~mag. To remove the Eddington bias from our samples, we do not admit single-epoch $W2$ measurements more than $1$~mag from the AllWISE catalog value.

We additionally performed two quality control cuts. First, a small percentage ($\sim0.9$) of our sample have large differences ($>0.5$~dex) between their S\'{e}rsic masses and their elliptical Petrosian masses, which can occur when the galaxy is cut off in its SDSS image. We removed these sources. Second, we cross-matched the AllWISE source coordinates of our sample to SDSS~DR12 to within $10\arcsec$ using \textsc{topcat} \citep{2005ASPC..347...29T},\footnote{\url{http://www.star.bris.ac.uk/~mbt/topcat}} version 4.7, returning the closest match for each source. We found that for sources with a larger NSA-AllWISE positional offset than their SDSS-AllWISE offset, removing objects with an NSA-SDSS offset greater than $1\arcsec$ cleanly removes spurious AllWISE counterparts. After performing these quality control cuts, we were left with 2199 dwarf galaxies and a pool of 31\,716 potential controls, from a starting sample of 2462 dwarf galaxies and 32\,577 potential controls.

The sensitivity of any metric to variability scales inversely with the signal-to-noise (S/N) of the data \citep[e.g.,][Figure~1]{2016ApJ...817..119K}, and so we produced a final sample of control galaxies matched to our dwarf galaxy sample in the following manner: for each dwarf galaxy, we determined the number of control galaxies available within some magnitude tolerance $\Delta W1$ such that every dwarf galaxy has the same number of controls and all controls are unique. We chose the $W1$ band because differences in $W1-W2$ for extragalactic objects are generally a result of brighter $W2$ relative to $W1$ (redder $W1-W2$), which does not bias against sensitivity to variability in $W2$. We found that choosing a tolerance of $\Delta W1\pm0.05$~mag produces a sample of 6591 controls matched to 2197 dwarf galaxies, where there are two fewer dwarf galaxies than we initially selected in Section~\ref{subsection: samples} because the bins are centered around the dwarf galaxy magnitudes and so the two faintest dwarf galaxies got cut. The mean $W1$ magnitude for the control and dwarf galaxies is $14.198\pm0.005$ mag and $14.198\pm0.008$ mag, and a Kolmogorov–Smirnov (K-S) test gives $p=1.0$, indicating that there is no bias in our variability metrics in the dwarf sample relative to the higher mass control galaxies caused by brightness differences.

Finally, as the dwarf galaxies and controls are separated in redshift by $\sim0.1$, we estimate their AllWISE magnitude K~corrections by doing SED decomposition using their Galactic extinction-corrected optical photometry from the NSA catalog and their AllWISE catalog photometry. We use the SED-fitting code employed in \citet{2018ApJ...858..110P} and \citet{2019ApJ...875..117P} with the AGN and galaxy templates from \citet{2010ApJ...713..970A}. The K~corrections are calculated as the difference between the observed-frame \textit{WISE} magnitudes of the rest-frame and redshifted best-fit linear combination of templates. These corrections are then applied to the observed catalog magnitudes. The AllWISE colors $W1-W2$ and $W2-W3$ given throughout this work are rest-frame.

\subsection{Variability Analysis} \label{subsection: analysis}
We explored several non-parametric variability metrics commonly employed in the literature. We calculate the variances, covariances, and Pearson $r$ as in \citet{2016ApJ...817..119K}, equations 1--3, the amplitude of variability following \citet{2007AJ....134.2236S}, equation 4 \citep[see also][]{2019MNRAS.483.2362R}, and the combined significance of variances in $W1$ and $W2$ following \citet[][Equation~4]{2016ApJ...817..119K}. To account for additional systematic uncertainty in the \textit{WISE} magnitudes, we add $0.024$~mag and $0.028$~mag in quadrature to the formal $W1$ and $W2$ band uncertainties \citep{2011ApJ...735..112J}. During our preliminary inspection of the light curves of objects exhibiting variability according to some of these metrics, we found photometric outliers in some of the single measurements that led to potential false positives. To address this, we exploited the fact that the single 7.7-second $W1$ and $W2$ measurements of a given source are strongly clustered in time because of the \textit{WISE} scanning pattern \citep[][Section~2.1]{2010AJ....140.1868W}. We binned single measurements into $\pm10$~day ``epochs'', and rejected outliers by removing measurements deviating by more than 3 standard deviations from the median magnitude of the epoch. To robustly estimate the standard deviation, we used the median absolute deviation \citep[e.g.,][]{1981rost.book.....H}, and the calculation of the median is weighted by the inverse variance of the data to account for differing measurement uncertainties. The total number of single measurements per epoch and the length of time between the first and last measurement depend on ecliptic latitude, but for our dwarf and control samples these are on average 11 measurements over 2 days,\footnote{The $5\sigma$ sensitivity of \textit{WISE} over 11 measurements is 17.11~mag in $W1$ and 15.66~mag in $W2$ \citep{2010AJ....140.1868W}. with a mean interval between measurements of 0.20 days. Fewer than $0.1\%$ of epochs are longer than 7.2 days, and the median time separating them is 188 days. The total duration of time between the first measurement in the first epoch to the last measurement in the last epoch is on average 3060 days, or 8.4 years.} We note that since the mid-infrared emission arises from spatial scales much larger than the optical and UV accretion disk emission, short term variability is expected to be ``smoothed out'' in the reprocessed mid-infrared emission from the dust, a result that is confirmed in previous mid-infrared variability studies \citep{2016ApJ...817..119K, 2010ApJ...716..530K,2019MNRAS.483.2362R}.

\begin{figure}
\center
  \includegraphics[width=\linewidth]{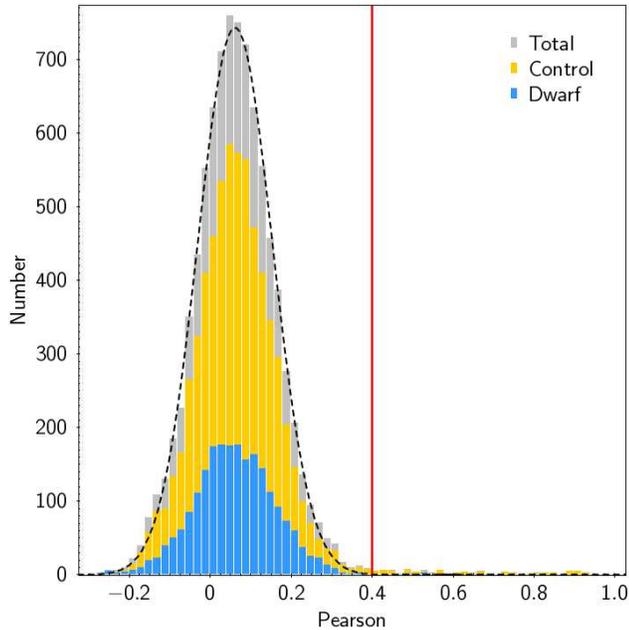}
  \caption{Distribution of the Pearson correlation coefficient between $W1$ and $W2$, with the sub-samples of dwarf galaxies and their controls labeled. Below a value of $+0.3$, the distribution is normally-distributed (dashed line). We choose a cut of Pearson $>+0.4$ (vertical line) to select candidate variables.}
  \label{fig: pearsoncut}
\end{figure}

With variability metrics calculated for the cleaned data, we found that the distribution of Pearson correlation coefficients $r$ between $W1$ and $W2$ follows a normal distribution below a value $+0.3$, with a mean of $+0.06$ and a standard deviation of $0.09$ (Figure~\ref{fig: pearsoncut}). After considering several variability metrics, we chose Pearson's $r$ to minimize false positives and produce a sample of candidate variables:

\begin{equation} \label{eq: pearson}
r = \frac{C_{m_1,m_2}}{\sigma_{m_1} \sigma_{m_2}},
\end{equation}

\noindent where $C_{m_1,m_2}$ is the covariance between the first and second bands (i.e., $W1$ and $W2$) for $N$ single measurements of an individual source:

\begin{equation} \label{eq: cov}
C_{m1,m2} = \frac{1}{N-1} \sum_{i}^{N} (m_{1,i} - \langle m_1 \rangle) \times (m_{2,i} - \langle m_2 \rangle),
\end{equation}

\noindent and $\sigma_{m_1}$, $\sigma_{m_2}$ are the variability amplitudes:

\begin{equation} \label{eq: var amp}
\sigma_m^2 = \frac{1}{N-1} \sum_{i}^{N} (m_i - \langle m \rangle)^2,
\end{equation}
\noindent where $m_i$ is the magnitude of the source during epoch $i$, and $\langle m \rangle$ is the mean magnitude of the source in the given band. We chose a cut of $r\geq+0.4$, which contains the majority of the non-Gaussian tail of potential variables (Figure~\ref{fig: pearsoncut}) and produced a sample of reasonable size for manual inspection. In manually inspecting the light curves, we randomly mixed the dwarf and control galaxy light curves together to avoid cognitive bias, and to err on the side of caution we did not include light curves with variability only between the AllWISE MEP data and the NEOWISE-R data in our final sample of bona fide variable light curves. While there are undoubtedly objects for which significant variability occurred only during the lacuna between the epochs of the AllWISE MEP and NEOWISE-R data, we required that there be variability within either the AllWISE MEP or NEOWISE-R data individually. 

\section{Results} \label{section: results}
Two of the 2197 dwarf galaxies in our sample are variable in the mid-infrared, or $0.09^{+0.20}_{-0.07}\%$ where the margins correspond to the 90\% confidence interval and are calculated using binomial statistics. By comparison, 66 of the 6591 more massive control galaxies are variable, or $1.00^{+0.23}_{-0.19}\%$, consistent with the findings of other mid-infrared variability studies targeting larger galaxies \cite[e.g.,][]{2010ApJ...716..530K, 2016ApJ...817..119K, 2018MNRAS.476.1111P}. The null hypothesis that the dwarf galaxies are as variable as these controls has a $p$-value of $6\times10^{-6}$. Larger galaxies with $M_\star\geq10^{10}$~$h^{-2}M_\sun$ therefore exhibit mid-infrared variability about ten times more frequently than dwarf galaxies with $M_\star<2\times10^{9}$~$h^{-2}M_\sun$. Within the control galaxy population, we find no evidence of a systematic difference in stellar mass between variable and non-variable sources, with mean stellar masses of $(3.47\pm0.19)\times10^{10}$~$h^{-2}M_\sun$ and $(3.82\pm0.03)\times10^{10}$~$h^{-2}M_\sun$, respectively. A two-sample K-S test between the stellar masses of variable and non-variable control galaxies gives $p=0.62$. This strongly suggests that the decrement of variable sources seen in the dwarf galaxy population occurs somewhere in the low-mass galaxy regime, between $2\times10^{9}$~$h^{-2}M_\sun$ and $10^{10}$~$h^{-2}M_\sun$.

\begin{figure*}
\center
  \includegraphics[width=\linewidth]{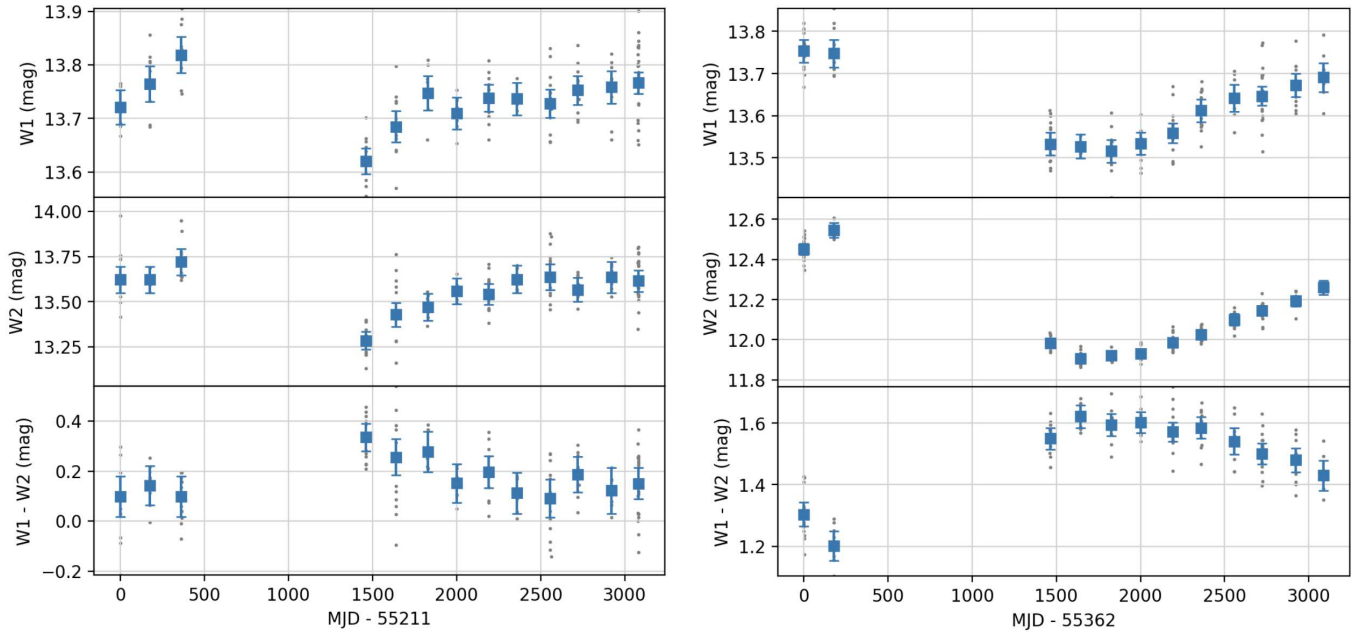}
  \caption{\textit{WISE} light curves of the two variable dwarf galaxies identified in this work. Gray points are the individual measurements; blue points show the weighted arithmetic mean and 2~times the corresponding standard error of the mean. \textit{Left:} NSAID 371667; \textit{Right:} NSAID 638093. Note the redder-when-brighter behavior, consistent with the brightening being due to increasing AGN dominance over the host galaxy.}
  \label{fig: dwarfvar}
\end{figure*}

Additionally, while we see a mild inverse correlation between stellar mass and variability amplitude $\sigma$ \citep[e.g.,][Equation~1]{2016ApJ...817..119K} in the controls, this correlation is very weak (Pearson~$r=-0.3$). The mean $\sigma_{W1}$, $\sigma_{W2}$ is $0.08\pm0.01$, $0.18\pm0.01$~mag for the dwarf galaxies and $0.11\pm0.01$, $0.14\pm0.01$~mag for the controls. The variability amplitude is therefore of similar magnitude for both the dwarf galaxies and their controls. We show the mid-infrared light curves for the two variable dwarf galaxies in Figure~\ref{fig: dwarfvar}. While only one galaxy has AGN-like $W1-W2$ color, both galaxies exhibit a redder-when-brighter trend, consistent with increasing AGN activity increasing the dominance of the AGN over its host galaxy.

\subsection{Relationship with Mid-Infrared Color} \label{subsection: mid-infrared color}
\begin{deluxetable*}{lrrccrr} \label{tab: colors}
\tablehead{
\colhead{} & \colhead{Total} & \colhead{                                                   } & \colhead{$g-r$} & \colhead{$W1-W2$} & \colhead{$W1-W2<0.5$} & \colhead{$W1-W2\geq0.8$} \\ [-0.5cm]
\colhead{} & \colhead{     } & \colhead{log$_{10}\left(\frac{h^{-2}M_\star}{M_\sun}\right)$} & \colhead{     } & \colhead{       } & \colhead{           } & \colhead{              } \\ [-0.4cm]
\colhead{} & \colhead{count} & \colhead{                                                   } & \colhead{ mag } & \colhead{  mag  } & \colhead{ count (\%)} & \colhead{    count (\%)}
}
\startdata
Dwarf & 2197 & $9.07\pm0.00$ & $0.478\pm0.003$ & $0.160\pm0.003$ & 2169 ($98.7^{+\hphantom{0}0.4}_{-\hphantom{0}0.5}$) & 10 ($0.46^{+0.32}_{-0.21}$) \\
Control & 6591 & $10.53\pm0.00$ & $0.681\pm0.001$ & $0.146\pm0.002$ & 6434 ($97.6^{+\hphantom{0}0.3}_{-\hphantom{0}0.3}$) & 40 ($0.61^{+0.18}_{-0.15}$) \\
Dwarf variable & 2 & $9.19\pm0.01$ & $0.517\pm0.060$ & $0.675\pm0.437$ & 1 ($50.0^{+47.5}_{-47.5}$) & 1 ($50.0^{+47.5}_{-47.5}$) \\
Control variable & 66 & $10.50\pm0.02$ & $0.610\pm0.016$ & $0.670\pm0.032$ & 18 ($27.3^{+10.4}_{-\hphantom{0}8.8}$) & 25 ($37.9^{+10.8}_{-10.0}$)
\enddata
\caption{Number of objects, mean stellar mass, and mean optical and mid-infrared color of the dwarf galaxies and matched controls examined in this work. Note that the error bounds on the percentage of objects with $W1-W2\geq0.8$ represent the 90\% confidence interval.}
\end{deluxetable*}

In Table~\ref{tab: colors}, we give the number of dwarf galaxies and controls with $W1-W2\geq0.8$~mag, where the mid-infrared color is generally dominated by the AGN \citep{2012ApJ...753...30S}. The percentage of all dwarf galaxies with $W1-W2\geq0.8$ is statistically indistinguishable from their controls ($p=0.30$). With only two variable dwarf galaxies, however, there is little constraint on what percentage of them have $W1-W2\geq0.8$, but for the controls this value is between 28\%--48\%. As these are 90\% confidence intervals, this additionally suggests that the reddest sources are not necessarily the most variable, and that mid-infrared variability serves as an additional method of AGN selection. Taken as a whole, of the 68 total variable objects between the dwarf galaxies and the more massive controls, 26 have $W1-W2\geq0.8$ ($38.2^{+10.7}_{-9.9}\%$), suggesting that mid-infrared color selection using the AllWISE catalog misses about half of AGN candidates selected using mid-infrared variability. Indeed, of the 68 variable objects, 19 have $W1-W2<0.5$ ($27.9^{+10.3}_{-8.8}\%$). These 19 objects have a mean $W1-W2$ that is $0.177\pm0.033$~mag redder than the non-variable population as a whole, suggesting \emph{some} level of AGN contribution to their \textit{WISE} colors. These results are consistent with the results from the Spitzer~Deep,~Wide-Field~Survey \citep[SDWFS;][]{2010ApJ...716..530K, 2016ApJ...817..119K}, where the mid-infrared variable sources are seen to extend blue-ward outside the mid-infrared AGN selection wedge from \citet{2005ApJ...631..163S}, uncovering a new population of AGNs that are undetected by a commonly employed mid-infrared color selection.

Although the true percentage of variable dwarf galaxies with $W1-W2\geq0.8$ cannot be determined with our sample, the percentage of dwarf galaxies with $W1-W2\geq0.8$ that show variability is better constrained. While 25 of the 40 control galaxies with $W1-W2\geq0.8$ are variable ($63^{+13}_{-14}\%$), only 1 of the 10 dwarf galaxies with $W1-W2\geq0.8$ is variable ($10^{+29}_{-9}$\%), a result that has a null hypothesis $p$-value of $p=0.008$. Thus, mid-infrared variability is identifying a smaller fraction of sources with red mid-infrared colors in the dwarf sample than in the more massive controls.

\begin{figure*}
\center
  \includegraphics[width=\linewidth]{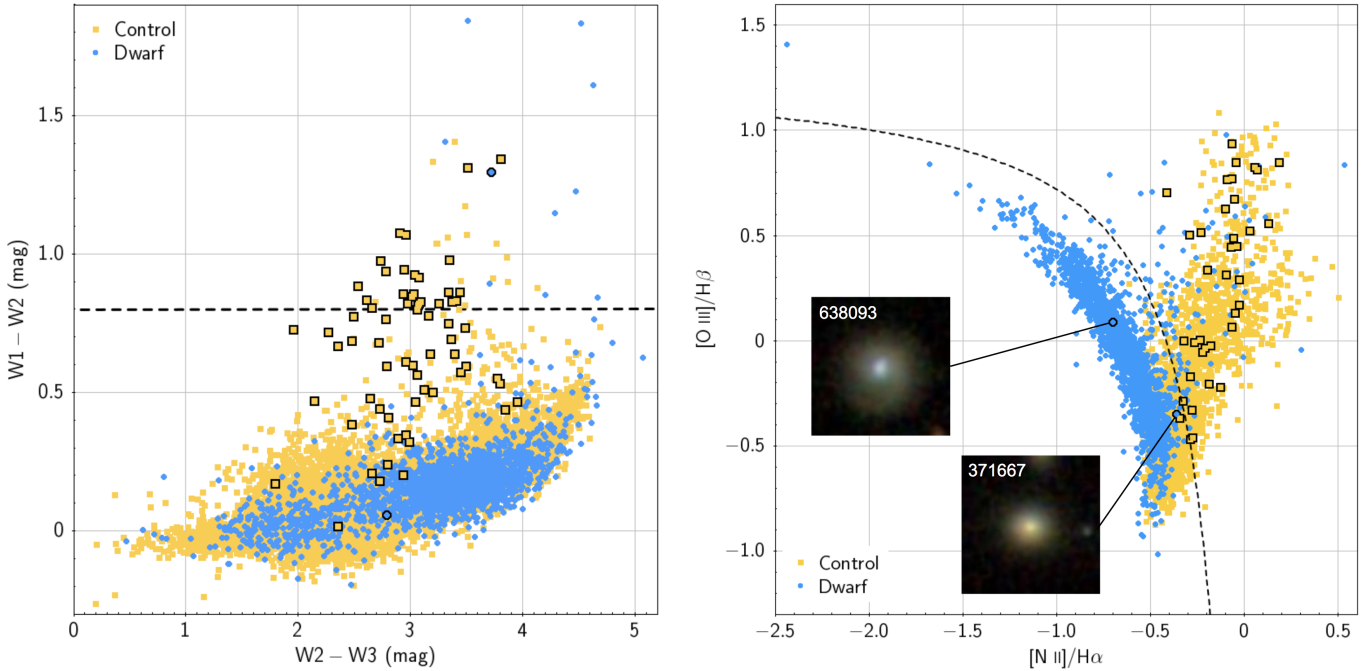}
  \caption{\textit{WISE} color-color diagram (left) and BPT diagram (right) of the dwarf galaxies and controls, with variable sources outlined. We plot the \citet{2012ApJ...753...30S} demarcation as a dashed line in the \textit{WISE} color-color diagram and the \citet{2003MNRAS.346.1055K} demarcation as a dashed line in the BPT diagram. For reference, we include $24\arcsec$ SDSS thumbnails of NSAID~638093 and NSAID~371667. Despite the extreme $W1-W2$ color of NSAID~638093, it is optically star-forming, while NSAID~371667, which is closest to being a BPT AGN, has \textit{WISE} colors consistent with a typical star-forming galaxy.}
  \label{fig: w123_bpt}
\end{figure*}

\subsection{Relationship with Optical Classification} \label{subsection: optical classification}
To explore the connection between optical emission line classification and mid-infrared variability, we cross-matched the NSA coordinates of our dwarf and control sample to the MPA-JHU catalog for SDSS~DR7,\footnote{\url{https://wwwmpa.mpa-garching.mpg.de/SDSS/DR7}} to within $1\farcs5$ to ensure that the measured object center of the galaxies falls within the $3\arcsec$~SDSS fiber. This catalog limits the Balmer line widths to $\leq500$~km~s$^{-1}$, and so does not contain many broad line objects. To recover these, we additionally matched our sample to the \texttt{specObj} table for SDSS~DR16\footnote{\url{https://www.sdss.org/dr16/spectro/spectro\_access}} to within $1\farcs5$ for the SDSS spectra and $1\arcsec$ for the BOSS spectra.\footnote{\url{https://www.sdss.org/dr16/spectro/spectro\_basics}} In all, of the 2197 dwarfs and 6591 controls, 8595 ($97.8\%$) have spectra from SDSS/BOSS covering their measured object centers. We classified the galaxies as either quiescent, star-forming, or AGN based on the following criteria:

\begin{enumerate}
    \item Quiescent: S/N~$<3$ for any of H$\beta$, [O\,\textsc{iii}]$\lambda5007$, H$\alpha$, or [N\,\textsc{ii}]$\lambda6584$. If no counterpart in the MPA catalog exists, then exclude any objects of \texttt{CLASS}==``QSO'', or \texttt{SUBCLASS} contains ``AGN'', ``STARFORMING'', or ``STARBURST''. 4791 objects meet these criteria (55.7\%).
    \item AGN: S/N~$\geq3$ in the four lines listed above and meeting the \citet{2003MNRAS.346.1055K} AGN criterion if in the MPA catalog, \texttt{CLASS}==``QSO'' or \texttt{SUBCLASS} contains ``AGN'' otherwise. 1121 objects meet this criteria (13.0\%).
    \item Star-forming (SF): not quiescent and not AGN. 2683 objects meet this criterion (31.2\%).
\end{enumerate}

\noindent The results of this classification are shown in Table~\ref{tab: spec}. Although both of the variable dwarf galaxies are emission line objects, neither of them are optically classified as AGNs, although as with mid-infrared color AGN selection this sets almost no constraint on the true fraction of variable dwarf galaxies that are optical AGNs. Nearly all of the variable control galaxies, however, are optical AGNs, a result that is highly significant ($p\sim0.0$). Comparing the dwarf galaxies to the controls, the probability of having no optical AGNs in a sample of two variable dwarf galaxies, given the frequency of optical AGNs in the variable control galaxies, is only $0.009$. We note that at the median redshift of the dwarf galaxies ($z=0.025$), a $3\arcsec$ spectroscopic fiber covers the inner $0.55$~$h^{-1}$~kpc of the nucleus, while at the median redshift of the controls ($z=0.135$) the fiber covers the inner $2.9$~$h^{-1}$~kpc, or about $\sim30$ times more area. Optical emission line dilution from star formation may therefore be more likely in the control galaxies for a given star formation rate, and so the true fraction of control galaxies with optical AGNs may be even higher.

\begin{deluxetable*}{lrrrr} \label{tab: spec}
\tablehead{
\colhead{}       & \colhead{Total} & \colhead{Quiescent}            & \colhead{SF}                   & \colhead{AGN}         \\ [-0.2cm]
\colhead{}       & \colhead{count} & \colhead{count (\%)}           & \colhead{count (\%)}           & \colhead{count (\%)}
}
\startdata
Dwarf            & 2068            & 286  ($13.8^{+1.3}_{-1.2}$)    & 1696 ($82.0^{+1.4}_{-1.4}$)    & 86   ($4.16^{+0.80}_{-0.70}$) \\
Control          & 6527            & 4505 ($69.0^{+0.9}_{-1.0}$)    & 987  ($15.1^{+0.7}_{-0.7}$)    & 1035 ($15.9^{+0.8}_{-0.7}$)   \\
Dwarf variable   & 2               & 0    ($0.0^{+77.6}_{-0.0}$)    & 2    ($100.0^{+0.0}_{-77.6}$)  & 0    ($0.0^{+77.6}_{-0.0}$)   \\
Control variable & 64              & 5    ($7.81^{+7.91}_{-4.68}$)  & 1    ($1.56^{+5.64}_{-1.48}$)  & 58   ($90.6^{+5.2}_{-8.3}$) 
\enddata
\caption{Optical spectroscopic properties of the dwarf and control galaxies examined in this work.}
\end{deluxetable*}

As was the case with mid-infrared selected AGNs in Section~\ref{subsection: mid-infrared color}, the fraction of optical AGNs in dwarf galaxies that exhibit variability is better constrained, with an upper limit of $<3.4\%$. This is statistically inconsistent ($p=0.01$) with the fraction of optical AGNs in the more massive control galaxies that exhibit variability ($5.6^{+1.3}_{-1.1}\%$), suggesting that while variability selection misses the vast majority of optically-selected AGNs in both samples, it is particularly biased against those in the dwarf galaxies.

\subsection{Comparison with Optical Variability-Selected AGNs in Dwarf/Low-Mass Galaxies} \label{subsec: comparison}
The results of this study demonstrate that the fraction of mid-infrared variable sources is significantly lower in dwarf galaxies than in a higher stellar mass sample of galaxies controlled for sensitivity to variability, appearing to suggest that there is a drop in the AGN occupation fraction at the lowest stellar masses. However, a comparison of AGN identification through optical emission line and mid-infrared color selection diagnostics reveals that AGNs selected using these diagnostics are less likely to be identified as variable in dwarf galaxies compared to their more massive controls. This suggests that variability selection is biased against finding AGNs in dwarf galaxies compared to their more massive counterparts, using data with a baseline and cadence similar to the \textit{WISE} observations.

To place these findings in the context of AGN candidates in dwarf or low-mass galaxies selected using variability at other wavelengths, we compare our results to those of \citet{2018ApJ...868..152B}, who looked for optical photometric variability in low mass galaxies selected from the NSA catalog that fall within SDSS Stripe~82. The baseline of their data was $\sim2300$ days, or 6.3 years, similar in scale to the \textit{WISE} data we employ here. The light curves produced by \citet{2018ApJ...868..152B} typically have about $\sim20$ individual photometric measurements each, but the majority of these data points are clustered with a few tens of days separating them, with the result that the 6.3-year light curves effectively have about 5--6 epochs \citep[see Figure~2 in][]{2018ApJ...868..152B}, giving a cadence of about $\sim400$ days, a factor of $\sim2$ less frequent than the WISE data excepting the lacuna between the AllWISE-MEP and NEOWISE-R data. \citet{2018ApJ...868..152B} find 135 AGN candidates out of 28\,062 galaxies (0.5\%), 35 of which are in low-mass ($M_\star<10^{10}$~$h^{-2}$~$M_\sun$) galaxies, and 12 dwarf galaxies out of 8941 in their parent sample are cited as having AGN-like variability. The percentage of dwarf galaxies in their sample that show optical variability is therefore $0.13^{+0.08}_{-0.06}\%$, where the uncertainties are again the 90\% confidence interval. Although the purpose of our work is not to perform an exhaustive search for all dwarf galaxies with mid-infrared variability, our sample of 2197 dwarf galaxies contains 2 objects that exhibit mid-infrared variability, or $0.09^{+0.20}_{-0.07}\%$, consistent with the optical variability fraction ($p=0.5$). While we do not have access to their exact numbers, Figure~6 in \citet{2018ApJ...868..152B} indicates that there are about 11\,000 galaxies with $M_\star>10^{10}$~$h^{-2}$~$M_\sun$ in their sample, 100 of which were found to be variable, or $0.91^{+0.16}_{-0.14}\%$. This is also statistically consistent with the value we find of $1.00^{+0.23}_{-0.19}\%$ ($p=0.33$), suggesting that optically-selected variability and mid~infrared-selected variability are similarly effective at finding AGN candidates in dwarf galaxies when using data with long baselines and sparse cadences, and that these candidates are about $\sim10$ times less frequent in dwarf galaxies than they are in larger galaxies.

Recently, \citet{2020ApJ...889..113M} pointed out that since the driving force of AGN variability likely arises from the accretion disk, variability timescales are expected to be significantly shorter for lower mass black holes expected in dwarf galaxies compared to more massive galaxies. Indeed, using the black hole to galaxy stellar mass relation for local AGNs from \citet{2015ApJ...813...82R}, who use $h=0.70$, the mean black hole mass of black holes in our dwarf galaxy sample is expected to be $\sim6\times10^{5}$~$M_\sun$, while for the larger controls the expected mean black hole mass is $2\times10^7$~$M_\sun$. Because of the low black hole mass expected in the dwarf galaxy population, \citet{2020ApJ...889..113M} point out that higher cadence observations are crucial for optical variability studies to be effective in finding lower mass black holes. Using data from HiTS, they find 502 variable sources in 12\,300 galaxies (4\%) with a mean stellar mass similar to the sample of \citet{2018ApJ...868..152B}. Both studies used $g$-band ($\sim0.5$\,\micron) data of galaxies selected from the SDSS, and both studies are sensitive to variability at about the $\sim3\%$ level at a $g$-band magnitude of $20$ \citep[see Figure~5 in][Section~5 in \citealt{2018AJ....156..186M}]{2018ApJ...868..152B}. While \citet{2020ApJ...889..113M} use apertures between $0\farcs6$--$2\farcs2$, smaller than the $2\farcs5$ aperture employed by \citet{2018ApJ...868..152B}, \citet{2020ApJ...889..113M} did not find any dependence between aperture size and variability, suggesting that choice of aperture is not a significant factor. \citet{2020ApJ...889..113M}, however, use data observed in $\sim2$~hr intervals over the span of 4--5 consecutive nights, giving a much shorter baseline and a much higher cadence. The difference in variability fraction of 4\% for the higher cadence HiTS data versus 0.5\% for the lower cadence Stripe-82 data suggest that higher cadence observations are about 8 times more sensitive to the presence of an AGN.

\section{Conclusions} \label{section: conclusions}
We have compared the mid-infrared photometric variability properties of a sample of dwarf galaxies ($M_\star<2\times10^{9}$~$h^{-2}M_\sun$) to a control sample, matched in mid-infrared apparent magnitude, of more massive galaxies ($M_\star\geq10^{10}$~$h^{-2}M_\sun$) drawn from the NSA catalog, version 1. Our conclusions are as follows:

\begin{enumerate}
    \item Of the 2197 dwarf galaxies in our sample, we find only 2 ($0.09^{+0.20}_{-0.07}\%$) with significant mid-infrared variability. By contrast, of the 6591 more massive control galaxies, 66 ($1.00^{+0.23}_{-0.19}$\%) show mid-infrared variability, in line with previous mid-infrared and optical variability studies in higher-mass galaxies. The null hypothesis that the dwarf galaxies and controls have the same frequency of mid-infrared variability has a $p$-value of $6\times10^{-6}$. We find no statistically significant difference between the stellar masses of the control galaxies that exhibit mid-infrared variability and the control galaxies that do not, suggesting that the sharp decline in apparent occupation fraction occurs between stellar masses of $2\times10^{9}$~$h^{-2}M_\sun$ and $10^{10}$~$h^{-2}M_\sun$.
    
    \item The amplitude of variability of the two dwarf galaxies is similar to that of the controls, with $\langle\sigma_{W1}\rangle=0.08\pm0.01$~mag, $\langle\sigma_{W2}\rangle=0.18\pm0.01$~mag for the dwarfs and $\langle\sigma_{W1}\rangle=0.11\pm0.01$~mag, $\langle\sigma_{W2}\rangle=0.14\pm0.01$ for the controls, consistent with previous variability studies of AGNs. We find only a very weak trend (Pearson~$r=-0.3$) of decreasing variability amplitude with increasing stellar mass for the controls. In both variable dwarf galaxies, brightening is associated with a reddening of their \textit{WISE} $W1-W2$ colors, consistent with increasing AGN dominance over the host galaxy.
    
    \item Of the two dwarf galaxies with mid-infrared variability, one displays AGN-like mid-infrared colors, but neither are optically classified as AGNs (either through their narrow emission line ratios or through the presence of broad lines). Only 1 of the 10 dwarf galaxies with AGN-like mid-infrared colors is variable, and none of the 86 dwarf galaxies classified as AGNs using optical emission line diagnostics is variable, results that are highly significant under the null hypothesis that the fraction of mid-infrared color-selected AGNs and optically classified AGNs exhibiting variability is the same as in the more massive control galaxies ($p=0.008$ and $p=0.01$, respectively). Optical spectroscopic and mid-infrared color selection is vastly more effective at finding AGNs in the low mass regime, underscoring the inefficiency of low cadence variability studies in finding AGNs in dwarf galaxies.
\end{enumerate}

\noindent Our results are consistent both with previous work exploring optical variability studies in low-mass and dwarf using data with a similar baseline and cadence, and with previous mid-infrared variability studies in larger galaxies. While mid-infrared variability may detect AGNs in dwarf galaxies not detected through other means, it is extremely inefficient, likely due to the baseline and cadence of the observations more than the wavelength. Shorter wavelength variability studies will likely benefit from higher-cadence observations when looking for AGNs in dwarf galaxies, which are expected to have variability timescales much shorter than in large galaxies owing to their less massive black holes.

\acknowledgements
We thank the anonymous referee for their helpful comments that greatly improved this work. This work also benefited from discussions with Matthew Ashby and Paulina Lira, who we gratefully acknowledge.

This research made use of Astropy,\footnote{\url{http://www.astropy.org}} a community-developed core Python package for Astronomy \citep{2013A&A...558A..33A, 2018AJ....156..123A}, as well as \textsc{topcat} \citep{2005ASPC..347...29T}.

Funding for the Sloan Digital Sky Survey IV has been provided by the Alfred P. Sloan Foundation, the U.S. Department of Energy Office of Science, and the Participating Institutions. SDSS-IV acknowledges support and resources from the Center for High-Performance Computing at the University of Utah. The SDSS web site is www.sdss.org.

SDSS-IV is managed by the Astrophysical Research Consortium for the Participating Institutions of the SDSS Collaboration including the Brazilian Participation Group, the Carnegie Institution for Science, Carnegie Mellon University, the Chilean Participation Group, the French Participation Group, Harvard-Smithsonian Center for Astrophysics, Instituto de Astrof\'isica de Canarias, The Johns Hopkins University, Kavli Institute for the Physics and Mathematics of the Universe (IPMU) / University of Tokyo, the Korean Participation Group, Lawrence Berkeley National Laboratory, Leibniz Institut f\"ur Astrophysik Potsdam (AIP), Max-Planck-Institut f\"ur Astronomie (MPIA Heidelberg), Max-Planck-Institut f\"ur Astrophysik (MPA Garching), Max-Planck-Institut f\"ur Extraterrestrische Physik (MPE), National Astronomical Observatories of China, New Mexico State University, New York University, University of Notre Dame, Observat\'ario Nacional / MCTI, The Ohio State University, Pennsylvania State University, Shanghai Astronomical Observatory, United Kingdom Participation Group, Universidad Nacional Aut\'onoma de M\'exico, University of Arizona, University of Colorado Boulder, University of Oxford, University of Portsmouth, University of Utah, University of Virginia, University of Washington, University of Wisconsin, Vanderbilt University, and Yale University.

This publication makes use of data products from the Wide-field Infrared Survey Explorer, which is a joint project of the University of California, Los Angeles, and the Jet Propulsion Laboratory/California Institute of Technology, and NEOWISE, which is a project of the Jet Propulsion Laboratory/California Institute of Technology. WISE and NEOWISE are funded by the National Aeronautics and Space Administration.

\facilities{NEOWISE, Sloan, WISE}

\software{
Astropy \citep{2013A&A...558A..33A,2018AJ....156..123A}, 
\textsc{topcat} \citep{2005ASPC..347...29T}
}

\bibliography{SecrestSatyapal20}{}
\bibliographystyle{aasjournal}

\end{document}